\documentclass[aps,nofootinbib,preprintnumbers,twocolumn,superscriptaddress]{revtex4-1}

\usepackage{amssymb,amsmath,bm,natbib}
\usepackage{color}
\usepackage{slashed}
\usepackage{graphics}
\usepackage{graphicx}
\usepackage[utf8]{inputenc}
\usepackage[caption=false]{subfig}
\usepackage[colorlinks = true,
            linkcolor = Maroon,
            urlcolor  = Maroon,
            citecolor = Maroon]{hyperref}
\usepackage{url}
\usepackage[dvipsnames]{xcolor}
\usepackage{dsfont}
\usepackage{float} 
\usepackage{cancel}
\usepackage{mhchem}
\usepackage{bm}
\usepackage[export]{adjustbox}
\usepackage{microtype}
\usepackage{soul}

\newcommand{\MeV}{{\rm Me\kern -0.1em V}}
\newcommand{\GeV}{{\rm Ge\kern -0.1em V}}
\newcommand{\TeV}{{\rm Te\kern -0.1em V}}
\newcommand{\eps}{\epsilon}

\newcommand{\e}{\ensuremath{\mathrm{e}}}

\usepackage{mfirstuc} 
\newcommand{\addReviewer}[2]{
  \expandafter\newcommand\csname #1\endcsname[1]{{\textbf{ \color{#2} \capitalisewords{#1}:\,##1}}}
  \expandafter\newcommand\csname #1cor\endcsname[2]{{\color{#2} \capitalisewords{#1}:\,\st{##1}{\textbf{##2}}}}
  \expandafter\newcommand\csname #1color\endcsname{#2}
  \expandafter\newcommand\csname #1todo\endcsname[1]{{\todo[inline,color=white!70!#2, caption={}]{\textbf{\capitalisewords{#1}}: ##1}}}
}

\usepackage{soul,color}
\definecolor{chromeyellow}{rgb}{1.0, 0.65, 0.0}

\addReviewer{ale}{chromeyellow}
\addReviewer{CAM}{blue}
\addReviewer{ang}{red}

\begin{document}
\preprint{ZU-TH 40/21, PSI-PR-21-20}

\title{Hunting for tetraquarks in ultra-pheripheral heavy ion collisions}

\author{Angelo Esposito}
\email{angeloesposito@ias.edu}
\affiliation{Theoretical Particle Physics Laboratory (LPTP), Institute of Physics, EPFL, 1015 Lausanne, Switzerland}
\affiliation{School of Natural Sciences, Institute for Advanced Study, Princeton, New Jersey 08540, USA}

\author{Claudio Andrea Manzari}
\email{claudioandrea.manzari@physik.uzh.ch}
\affiliation{Physik-Institut, Universit\"at Z\"urich, Winterthurerstrasse 190, CH-8057 Z\"urich, Switzerland}
\affiliation{Paul Scherrer Institut, CH-5232 Villigen PSI, Switzerland}

\author{Alessandro Pilloni}
\affiliation{INFN Sezione di Roma, I-00185, Rome, Italy}
\affiliation{Dipartimento di Scienze Matematiche e Informatiche, Scienze Fisiche e Scienze della Terra, 
Universit\`a degli Studi di Messina, I-98122 Messina, Italy}
\affiliation{INFN Sezione di Catania, I-95123 Catania, Italy}

\author{Antonio Davide Polosa}
\affiliation{INFN Sezione di Roma, I-00185, Rome, Italy}
\affiliation{Dipartimento di Fisica, Sapienza Universit\`a di Roma, I-00185 Roma, Italy}


\begin{abstract}
Ultra-peripheral heavy ion collisions constitute an ideal setup to look for exotic hadrons because of their low event multiplicity and the possibility of an efficient background rejection. We propose to look for four-quark states produced by photon-photon fusion in these collisions at the center-of-mass energy per nucleon pair $\sqrt{s_\text{NN}}=5.5~\TeV$. In particular, we focus on those states that would represent a definite smoking gun for the compact tetraquark model. We show that the $X(6900)$, a likely $cc\bar{c}\bar{c}$ compact state, is a perfect candidate for this search, and estimate a production cross section ranging from around $250$~nb to $1150$~nb, depending on its  quantum numbers. Furthermore, we discuss the importance of ultra-peripheral collisions to the search for the scalar and tensor partners of the $X(3872)$ predicted by the diquarkonium model, and not yet observed. The completion of such a flavor-spin multiplet would speak strongly in favor of the compact tetraquark model.
\end{abstract}
\maketitle


\section{Introduction}

The existence of hadrons with more than three valence constituents is now well assessed~\cite{Esposito:2016noz,Guo:2017jvc,Olsen:2017bmm,Brambilla:2019esw}, but the understanding of their nature remains a long standing problem of low-energy QCD. Are these states extended hadronic molecules arising from color neutral interactions? Or are they rather compact tetraquarks generated by short distance forces, analogs to mesons and baryons?

The solution to this issue requires the identification of some smoking guns, able to clearly discriminate between the two models. One such possibility is the recent observation by LHCb of a narrow resonance in the di-$J/\psi$ mass spectrum~\cite{Aaij:2020fnh}, dubbed $X(6900)$ and compatible with a $cc\bar c\bar c$ structure. The possibility of such a state was already anticipated by several studies, and later further investigated (see, e.g.,~\cite{Chao:1980dv,Heller:1985cb,Barnea:2006sd,Vijande:2007ix,Ebert:2007rn,Berezhnoy:2011xn,Wu:2016vtq,Chen:2016jxd,Wang:2017jtz,Debastiani:2017msn,Richard:2017vry,Anwar:2017toa,Bedolla:2019zwg,Karliner:2016zzc,Becchi:2020uvq,Dong:2020nwy,Cao:2020gul,Liang:2021fzr,Wang:2020wrp,Dong:2021lkh}). Crucially, no single light hadron  can mediate the interaction between charmonia to generate a loosely bound molecule~\cite{Maiani:2020pur,Dong:2021lkh}.
The $X(6900)$ seems likely to be a compact tetraquark.

Another compelling indication of the tetraquark nature of the exotic states would be the observation of a complete flavor-spin multiplet, as predicted in~\cite{Maiani:2014aja}. In the hidden charm sector, the $J^{PC}=1^{+-}$ resonances---the so-called $Z_c(3900)$ and $Z_c^\prime(4020)$---have been observed in three charge states, while the $1^{++}$ one---the famous $X(3872)$---has only been observed in a single neutral component. Besides the charged partners of the $X(3872)$, to complete the $c \bar c q \bar q$ multiplet, one would have to observe the predicted scalar and tensor states~\cite{Maiani:2014aja}.

In this work we propose to look for the above-mentioned smoking guns in ultra-peripheral heavy ion collisions (UPCs) at the LHC. In these events the impact parameter is much larger than the ions' radii, which then scatter off each other elastically~\cite{Baur:2001jj,Bertulani:2005ru,Baltz:2007kq}. This causes a lack of additional calorimetric signals and a large rapidity gap between the particles produced and the outgoing beams, which can be used for an efficient background rejection. For this reason they are an optimal environment for exotic searches, ranging from hadronic states to extra dimensions (see, e.g.,~\cite{Grabiak:1987uf,Drees:1989vq,Greiner:1992fz,Ahern:2000jn,Bertulani:2009qj,Goncalves:2012zm,Moreira:2016ciu,Knapen:2016moh,Goncalves:2018hiw,Goncalves:2021ytq}). These collisions are particularly amenable to search for states, like the ones of interest to us, that can be produced by photon-photon fusion. Indeed, the large charge of lead ions ($Z=82$) induces a huge $Z^4$ enhancement in the coherent photon--photon luminosity, consequently boosting the production cross section for these states.

The results we find are very encouraging. At the center-of-mass energy per nucleon pair $\sqrt{s_\text{NN}}=5.5~\TeV$, both the $X(6900)$ and the scalar and tensor $c\bar c q\bar q$ states are expected to be copiously produced in UPCs. In particular, due to its likely large width into vector charmonia, the $X(6900)$ should be produced with cross sections of the order of fractions of microbarn, or even more. The scalar and tensor states of the $X(3872)$ multiplet should instead be produced with cross sections larger than the measured one of the $X(3872)$ in prompt $pp$ collisions~\cite{CMS:2013fpt,LHCb:2021ten}. The observation of these states in UPCs would be another indication of the existence of compact tetraquarks in the spectrum of short distance QCD, alongside with a recently emerging pattern which includes the observation of the hidden charm and strange states~\cite{BESIII:2020qkh,LHCb:2021uow,Maiani:2021tri} and the study of the lineshape of the $X(3872)$~\cite{LHCb:2020xds,Esposito:2021vhu}.


\section{Photon--photon interaction}

When two ions pass each other at distances larger than their radii they interact solely via their electromagnetic fields.
For relativistic ions with $Z\gg 1$, the electric and magnetic fields are perpendicular, and the configuration may be represented as a flux of almost-real photons following the Wizs\"acker--Williams method~\cite{vonWeizsacker:1934nji,Williams:1934ad}. In particular, the number of photons per unit area and energy emitted by an ion with boost factor $\gamma\gg 1$ is given by~\cite{Klein:2016yzr}
\begin{align}
    \frac{dN_\gamma(k,\bm{b})}{dk \, d^2b} = \frac{Z^2\alpha}{\pi^2}\frac{k}{\gamma^2}K_1^2\left(\frac{k|\bm b|}{\gamma}\right)\,,
    \label{eq:PhotonDensity}
\end{align}
where $k$ is the photon energy, $\bm b$ is the transverse distance from the moving ion, $\alpha$ is the electromagnetic fine structure constant and $K_1$ is the modified Bessel functions. Since the photons are quasireal, in Eq.~\eqref{eq:PhotonDensity} only the flux of transversely polarized photons has been considered.

In an UPC the two-photon luminosity is given by
\begin{align}
\begin{split}
    \frac{d^2N_{\gamma\gamma}(k_1,k_2)}{dk_1dk_2} = & \int d^2b_1d^2b_2 \, P_{\rm NOHAD}\big(|\bm{b}_1-\bm{b}_2|\big) \\
    & \times \frac{dN_\gamma(k_1,\bm{b}_1)}{dk_1 \, d^2b_1}\frac{dN_\gamma(k_2,\bm{b}_2)}{dk_2 \, d^2b_2} \,,
    \end{split}
\end{align}
which evidently features a $Z^4$ enhancement---see Eq.~\eqref{eq:PhotonDensity}.
The requirement that the two nuclei do not interact hadronically is imposed by $P_{\rm NOHAD}(b)$, which is the probability of having no hadronic interactions at impact parameter $b$. In what follows we use the \texttt{STARlight} code~\cite{Klein:2016yzr}, where 
\begin{align}
    P_{\rm NOHAD}(b) = \e^{-\sigma_{\rm NN}T_{\rm AA}(b)}\,,
\end{align}
with $\sigma_{\rm NN}$ the nucleon--nucleon interaction cross section, and $T_{\rm AA}(b)=\int d^2b_1\, T_A(b_1)\, T_A(|\bm{b}_1-\bm{b}|)$ the nuclear overlap function determined from the Woods--Saxon nuclear density distributions of the two nuclei, $T_A(b)$.

We are interested in processes where the two photons produce a state, $X$, with invariant mass $W=\sqrt{4k_1k_2}$ and rapidity $Y=\frac{1}{2}\ln(k_1/k_2)$. The cross section for such a process factorizes in two terms: the two-photon luminosity associated to the incoming nuclei, and the cross section, $\sigma_{\gamma\gamma}(W)$, for the creation of $X$ from two photons, i.e.
\begin{align}
    \sigma(\ce{Pb}\,\ce{Pb} \to \ce{Pb}\,\ce{Pb}\, X) = \int dYdW \frac{d^2N_{\gamma\gamma}}{dWdY}\,\sigma_{\gamma\gamma}(W)\,.
\end{align}
The cross section to produce a single meson in a photon--photon interaction is given by~\cite{Klein:2016yzr}
\begin{align}
    \begin{split}
    \sigma_{\gamma\gamma}(W) &= 8\pi(2J+1)\frac{\Gamma_{\gamma\gamma}\Gamma}{(W^2-m^2)^2+m^2\Gamma^2}\\
    &\simeq 8\pi^2(2J+1)\frac{\Gamma_{\gamma\gamma}}{2m^2}\delta(W-m)\,,
    \end{split}
    \label{sigmagg}
\end{align}
where $m$ is the meson mass, $\Gamma_{\gamma\gamma}$ is its width in two photons, $\Gamma$ is the total width, $J$ is its spin. The last step realizes the narrow width approximation. From here we see that lighter and higher spin particles are produced more copiously.


\subsection{Partial widths into {\normalsize $\gamma\gamma$}}

\begin{figure}
    \centering
    \includegraphics[scale=0.7]{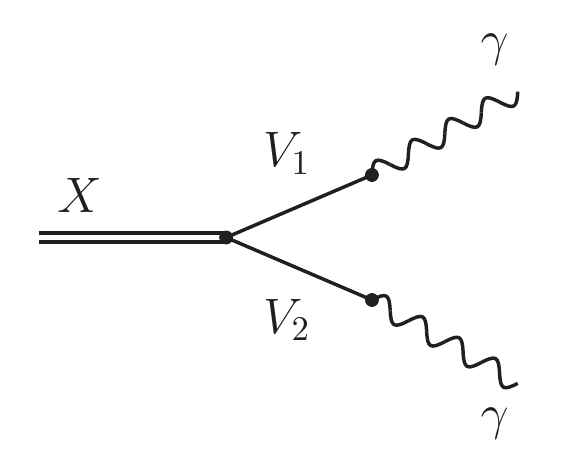}
    \caption{Decay of the state $X$ into $\gamma\gamma$ via two vectors.}
    \label{fig:diagram}
\end{figure}

It is clear that the central quantity in this formalism is the partial width of the state $X$ in two photons. In what follows we will compute it using the vector meson dominance model~\cite{SAKURAI19601}. In this picture, the radiative decay of a hadron happens first via its decay into vector mesons, which then mix with photons---see Figure~\ref{fig:diagram}.
\begin{table}[t]
    \centering
    \begin{tabular}{c||c|c|c}
         & $\rho$ & $\omega$ & $J/\psi$  \\
        \hline\hline
        $\Gamma_{ee}$ ($\GeV$)\; & \;$\simeq 7.0 \times 10^{-6}$\; & \;$\simeq 6.4 \times 10^{-7}$\; & \;$\simeq 5.5 \times 10^{-6}$\; \\
        \hline
        $f_V$ ($\GeV^2$) \; & \;$\simeq0.16$\; & \;$\simeq0.17$\; & \;$\simeq1.3$\;
    \end{tabular}
    \caption{Electronic widths as taken from PDG~\cite{ParticleDataGroup:2020ssz} and the decay constants extracted from them. The results are consistent with what was found in~\cite{Casalbuoni:1996pg,Deandrea:2002kj}.}
    \label{tab:fs}
\end{table}
In particular, the vector--photon mixing is given by
\begin{align*}
    \includegraphics[width=0.3\textwidth,valign=c]{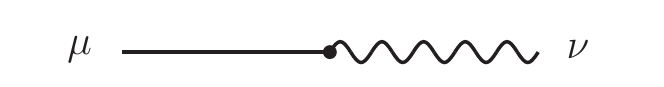} \hspace{-1.5em} &= \kappa_V\, e\, f_V\, \eta^{\mu\nu}\,,
\end{align*}
with $\kappa_V=\big(\frac{1}{\sqrt{2}},\frac{1}{3\sqrt{2}},\frac{2}{3}\big)$ if $V=(\rho,\omega,J/\psi)$.\footnote{
The vector-photon mixing is obtained from the standard electromagnetic Lagrangian, $\mathcal{L} = A_\mu \sum_a Q_a e \bar q_a \gamma^\mu q_a$, together with the meson states $|\rho \rangle = \frac{|u\bar{u}\rangle - |d\bar{d}\rangle}{\sqrt{2}}$, $|\omega \rangle = \frac{|u\bar{u}\rangle + |d\bar{d}\rangle}{\sqrt{2}}$ and $|J/\psi\rangle = |c\bar {c}\rangle$. The decay constants are defined through the matrix element $\langle 0 | \sum_a Q_a e \bar q_a \gamma^\mu q_a | V \rangle = \kappa_V f_V \epsilon^\mu$.}
The decay constants $f_V$ can instead be extracted from the electronic width of the corresponding vector, $\Gamma(V\to e^+e^-) = 4\pi \alpha^2 \kappa_V^2 f_V^2 / (3m_V^3)$. In Table~\ref{tab:fs} we report the electronic widths and the corresponding mixing constants.

The most general matrix elements for the decay of the scalar and tensor exotic mesons in two vectors can be written as
\begin{subequations} \label{eq:matrixelements}
\begin{align}
    \langle X(0^{++})|V_1 V_2\rangle &= \alpha_0\, \epsilon_{1}\cdot\epsilon_{2}  + \beta_0 \left( \epsilon_1\cdot k_2 \right)\left( \epsilon_2\cdot k_1 \right)\,, \\
    \langle X(2^{++}) | V_1 V_2 \rangle & = \pi_{\mu\nu} \big[ \alpha_2 \epsilon_1^\mu \epsilon_2^\nu + \beta_2 \left( \epsilon_1\cdot k_2 \right) \epsilon_2^\mu k_1^\nu \\
    & + \beta_2^\prime \left( \epsilon_2\cdot k_1\right) \epsilon_1^\mu k_2^\nu + \gamma_2 \left( \epsilon_1\cdot\epsilon_2 \right) k_1^\mu k_2^\nu \notag \\
    & + \delta_2 \left( \epsilon_1\cdot k_2 \right)\left( \epsilon_2\cdot k_1 \right) k_1^\mu k_2^\nu \big] \,, \notag
\end{align}
\end{subequations}
where $\epsilon_i$ and $k_i$ are the polarization and momentum of the vector $V_i$, and $\pi_{\mu\nu}$ is the polarization of the tensor.\footnote{The sum over spin-2 polarizations is given by~\cite{Faccini:2012zv,Gleisberg:2003ue} $\sum_{\text{pol}} \pi_{\mu\nu}(k)\pi_{\rho\sigma}(k)=\frac{1}{2}\mathcal{P}_{\mu\rho}\mathcal{P}_{\nu\sigma}+\frac{1}{2}\mathcal{P}_{\mu\sigma}\mathcal{P}_{\nu\rho}-\frac{1}{3}\mathcal{P}_{\mu\nu}\mathcal{P}_{\rho\sigma}$, with $\mathcal{P}_{\mu\nu}=-\eta_{\mu\nu}+k_\mu k_\nu/m^2$ and $k^2=m^2$.}
In absence of further information it is impossible to determine all the above couplings from the data. We will therefore adopt a minimal model, somewhat inspired from an EFT approach, and neglect all terms proportional to the particle momenta (see, e.g.,~\cite{Mathieu:2020zpm}).\footnote{For the scalar case, we checked that including the $\beta_0$ coefficient and letting it vary around its natural value, $\beta_0\sim 1/m_X$, does not change the order of magnitude estimates of Table~\ref{tab:sigmas6900}.}


\subsection{Production of the $X(6900)$}

As already mentioned, the $X(6900)$ is, in all likelihood, a compact $cc\bar c \bar c$ state. Its mass and width are $m_X=6886\pm 2~\MeV$ and $\Gamma_X=168\pm102~\MeV$~\cite{Aaij:2020fnh}, while its quantum numbers are yet to be determined. Were it to have $J^{PC}=0^{++}$ or $2^{++}$, it could be produced from photon--photon fusion in UPCs, as also discussed in~\cite{Goncalves:2021ytq}.

To provide an order of magnitude estimate of its partial width in two photons we make the assumption that its coupling to vector mesons is dominated by the di-$J/\psi$ one~\cite{Karliner:2016zzc}. Indeed, with four heavy quarks involved, the coupling to light vector mesons involves annihilation processes, and are thus OZI-suppressed by powers of $\alpha_s(4m_c)$. The contribution to the vector meson dominance from excited charmonia is also suppressed by their greater spatial extent~\cite{Redlich:2000cb}.\footnote{In~\cite{Goncalves:2021ytq} the partial width of the $X(6900)$ in two photons is taken to be the same as the $\chi_{cJ}$ quarkonium with the same quantum numbers. This underlines the somewhat strong assumption that the short distance dynamics of the two states is the same, which is not guaranteed. Here we take a more conservative approach and keep the branching ratio unspecified.}

Starting from the matrix elements in Eqs.~\eqref{eq:matrixelements}, and using the vector meson dominance as in Figure~\ref{fig:diagram}, we can obtain the partial width in two photons. Note that, since the amplitude is not gauge invariant, one must restrict oneself to the transverse photon polarizations. The results for the scalar and tensor case are
\begin{subequations}
\begin{align}
    \Gamma_{\gamma\gamma}^{0^{++}} &= \frac{16\pi \alpha^2}{81} \frac{\alpha_0^2}{m_X} \left( \frac{f_{\psi}}{m^2_{\psi}} \right)^4\,, \\
    \Gamma_{\gamma\gamma}^{2^{++}} & = \frac{56\pi \alpha^2}{1215} \frac{\alpha_2^2}{m_X}\left( \frac{f_{\psi}}{m_{\psi}^2} \right)^4\,.
\end{align}
\end{subequations}

The couplings $\alpha_J$ can be extracted from the partial width of the $X(6900)$ in di-$J/\psi$. Since the corresponding branching ratio is yet unknown, we will keep it general, bearing in mind that it is likely that this channel will dominate the total width~\cite{Karliner:2016zzc}. For the scalar and tensor cases one gets, respectively
\begin{subequations}
\begin{align}
    \mathcal{B}_{\psi}\, \Gamma_X &= \frac{\alpha_0^2 \, p}{16\pi m^2_X} \left( 3- \frac{m_X^2}{m_{\psi}^2} + \frac{1}{4} \frac{m_X^4}{m_{\psi}^4} \right)\,,
\end{align}
and
\begin{align}
    \mathcal{B}_{\psi}\,\Gamma_X &= \frac{\alpha_2^2 \, p}{16\pi m_X^2}\left( \frac{7}{15}+\frac{1}{10} \frac{m_X^2}{m_{\psi}^2}+\frac{1}{120}\frac{m_X^4}{m_{\psi}^4}\right)\,,
\end{align}
\end{subequations}
where $\mathcal{B}_{\psi}$ is the branching ratio of the di-$J/\psi$ final state, and $p=\lambda^{1/2}(m_X^2,m_{\psi}^2,m_{\psi}^2)/(2m_X)$ the decay momentum, with $\lambda$ the K\"all\'en function.

\begin{figure*}
    \centering
    \includegraphics[width=0.32\textwidth]{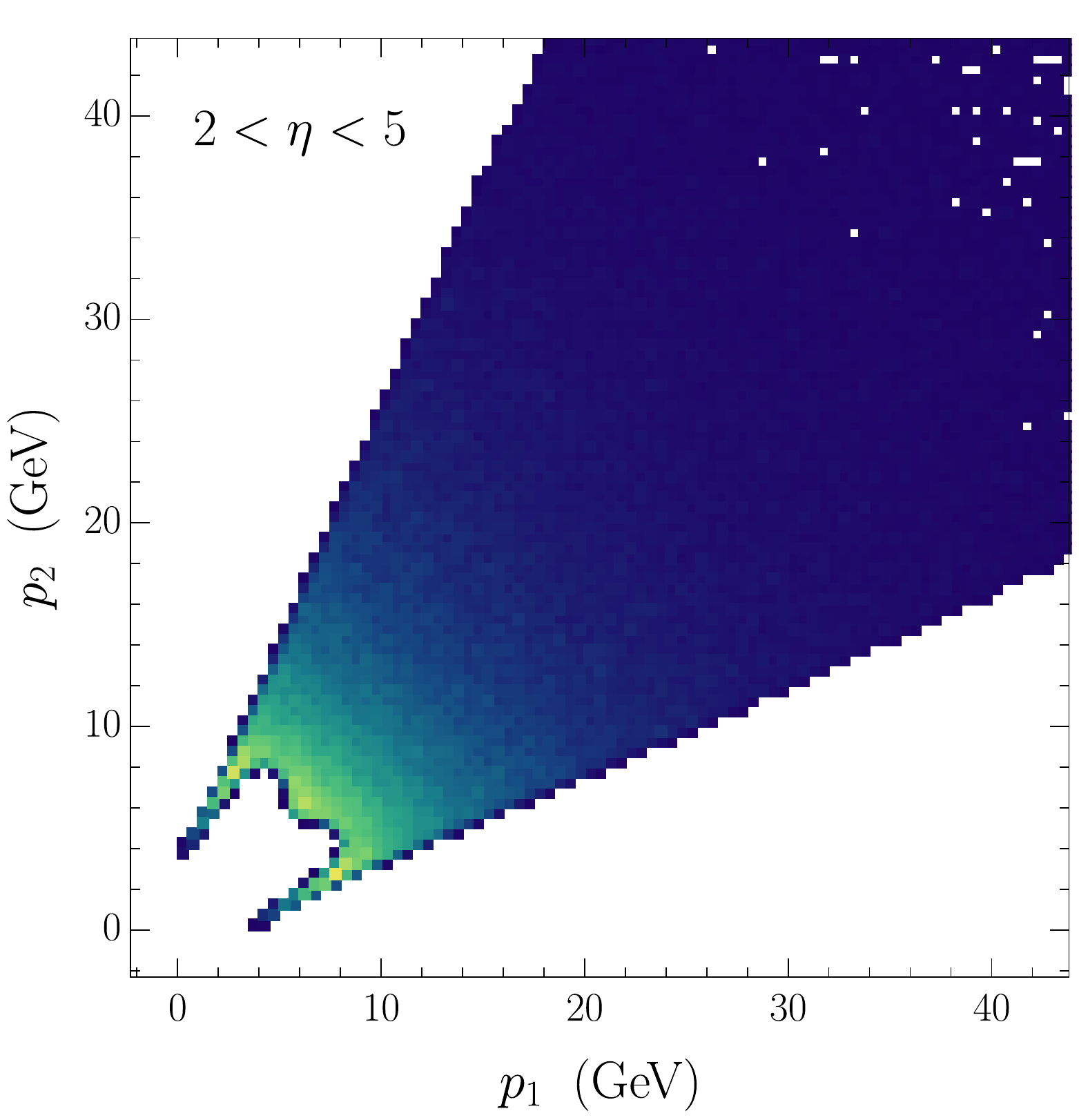}
    \includegraphics[width=0.325\textwidth]{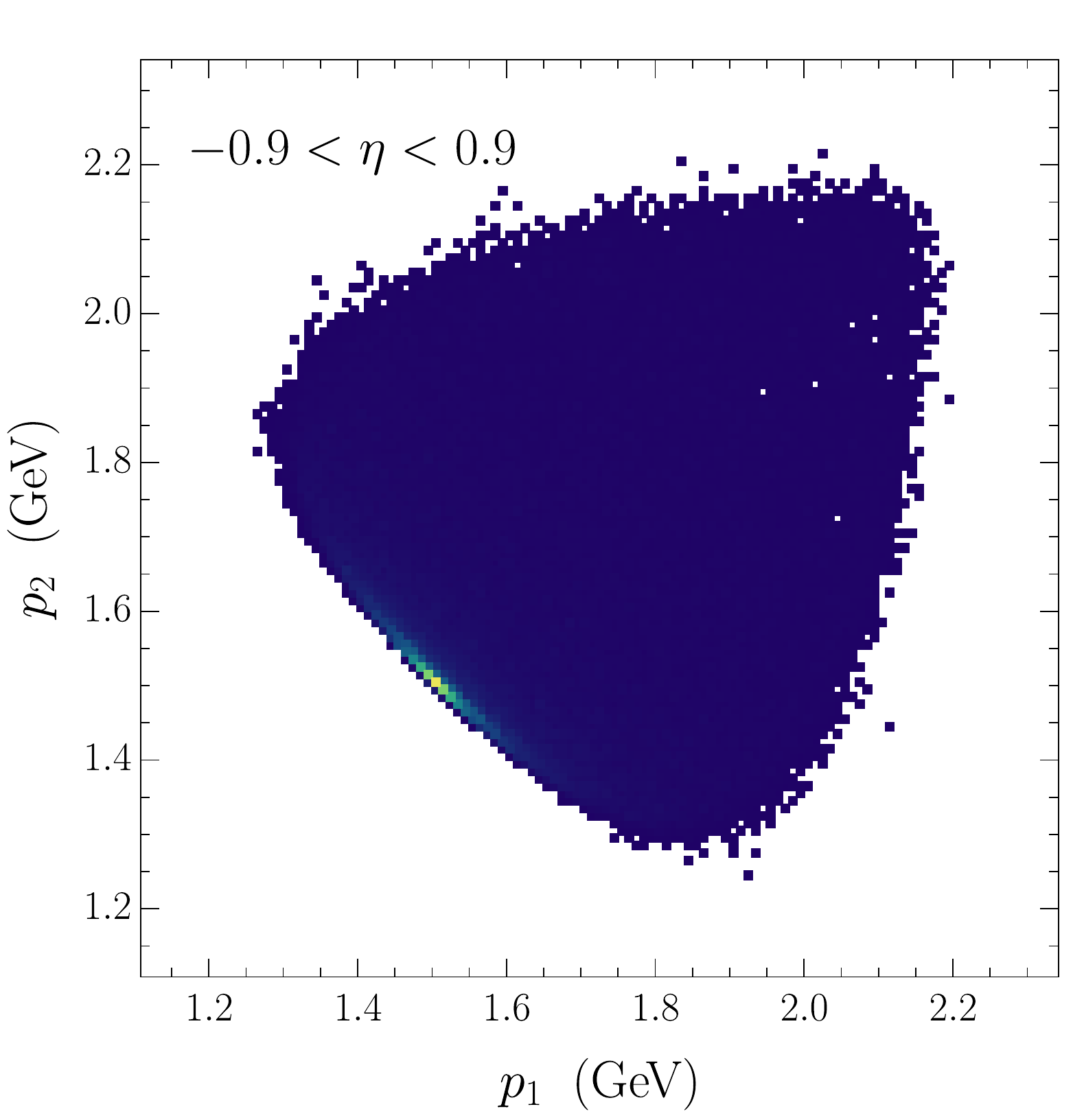}
    \includegraphics[width=0.32\textwidth]{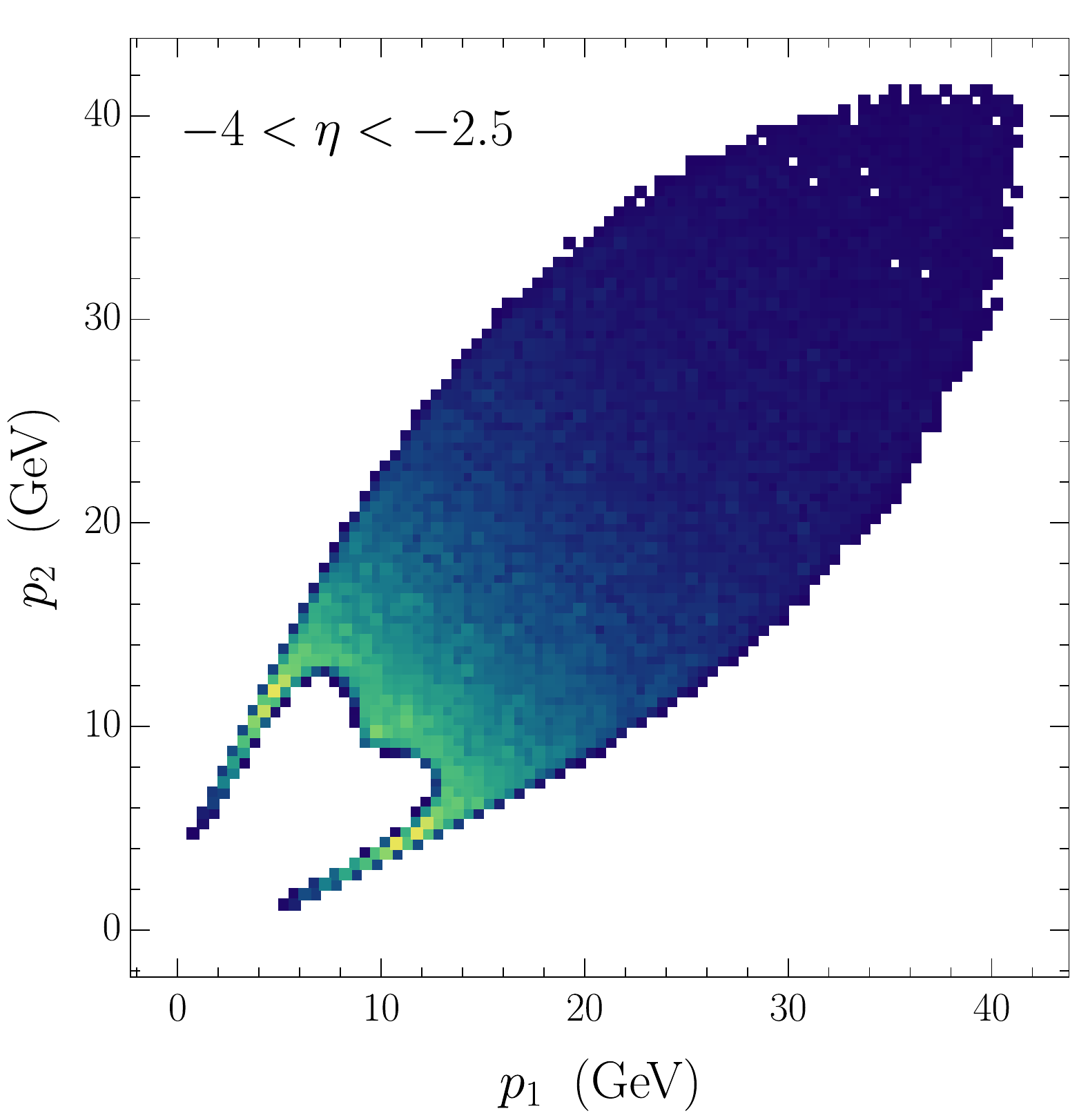}
    \caption{Momentum distributions of the two $J/\psi$'s produced by the decay of the $X(6900)$, for the pseudorapidity cuts corresponding to the LHCb detector (left), the ALICE electron detector (center) and the ALICE muon detector (right). While for the ALICE electron detector the two decay products are rather soft, the forward detectors should collect energetic ones. The pseudorapidity cuts have efficiencies of 21\%, 8\%, and 10\%, respectively.}
    \label{fig:dists}
\end{figure*}

In Table~\ref{tab:sigmas6900} we report the partial widths in two photons and the corresponding cross sections for production in UPCs as obtained from the \texttt{STARlight} code~\cite{Klein:2016yzr}.
\begin{table}[]
    \centering
    \begin{tabular}{c||c|c}
        State, $J^{PC}$ & $\Gamma_{\gamma\gamma}/\mathcal{B}_\psi$ (eV) & 
        $\sigma(\text{Pb}\text{Pb}\to\text{Pb}\text{Pb}X)/\mathcal{B}_{\psi}$ (nb) \\
        \hline\hline
        $X(6900)$, $0^{++}$ & $\sim 104$ & $\sim 282$\\
        $X(6900)$, $2^{++}$ & $\sim 86$  & $\sim 1165$
    \end{tabular}
    \caption{Partial widths in two photons and corresponding production cross sections in UPCs for the $X(6900)$, obtained for $\sqrt{s_{\text{NN}}}=5.5~\TeV$ and normalized by the di-$J/\psi$ braching ratio. The latter is unknown, but expected to be close to one.}
    \label{tab:sigmas6900}
\end{table}
In Figure~\ref{fig:dists} we report the momentum distributions of the two $J/\psi$'s produced by the decay of the $X(6900)$. We apply the pseudorapidity cuts corresponding to the LHCb and ALICE acceptances. As one can see, both experiments should be sensitive to energetic final states.


\subsection{Production of scalar and tensor $c\bar c q\bar q$ states}

Contrary to the $X(6900)$, the scalar and tensor states of the $c\bar c q \bar q$ multiplet are yet to be observed. The diquarkonium model predicts two $J^{PC}=0^{++}$ states, dubbed $X_0$ and $X_0^\prime$ and with masses around $m_0 \simeq 3770~\MeV$ and $m_{0^\prime}\simeq 4000~\MeV$, respectively~\cite{Maiani:2014aja}. It also predicts one $2^{++}$ state, dubbed $X_2$, and is degenerate with the $X_0^\prime$, $m_2 \simeq 4000~\MeV$. They all have the right quantum numbers to be produced via photon--photon fusion in UPCs.

As for the $X(3872)$, the states above can in principle decay into both $J/\psi\,\rho$ and $J/\psi\,\omega$. 
Indeed, the isospin breaking mechanism for tetraquarks holds regardless of the mass splitting in the multiplet~\cite{Rossi:2004yr}. We therefore assume that the scalars and tensor share the same isospin breaking pattern as the $X(3872)$.
In terms of the spins of the $c \bar c$ and $q\bar q$ pairs, one has~\cite{Maiani:2014aja}
\begin{subequations}
\begin{align}
    &|X_0\rangle = \frac{1}{2} |0_{c\bar c},0_{q\bar q}\rangle_0 + \frac{\sqrt{3}}{2}|1_{c\bar c},1_{q\bar q}\rangle_0 \,, \\
    &|X_0^\prime\rangle = \frac{\sqrt{3}}{2} |0_{c\bar c},0_{q\bar q}\rangle_0 - \frac{1}{2}|1_{c\bar c},1_{q\bar q}\rangle_0 \,, \\
    &|X_1\rangle \equiv |X(3872)\rangle  = |1_{c\bar c},1_{q\bar q}\rangle_1\,, \\[4pt]
    &|X_2\rangle = |1_{c\bar c}, 1_{q\bar q}\rangle_2\,,
\end{align}
\end{subequations}
where $|J_{c \bar c},J_{q\bar q}\rangle_J$ is a state with spin $J_{q_1\bar q_2}$ for the quark-antiquark pairs and total spin $J$. Considering again the matrix elements in Eqs.~\eqref{eq:matrixelements} under the minimal model,\footnote{In the case of the $X(3872)$, it has been shown~\cite{Maiani:2004vq} 
that the minimal model is able to reproduce the observed partial widths in $J/\psi\rho$ and $J/\psi\omega$. The dynamics of the members of the same multiplet will likely be similar, hence justifying restricting oneself to the minimal model for the scalar and tensor states as well.} one can relate their couplings to that of the $X(3872)$. The exact relations between the couplings to $J/\psi V$ ($V=\rho,\omega$) of the $X_0$, $X_0^\prime$, $X_2$, and $X(3872)$ depend on the dynamics of the multiplet. Being this level of precision negligible for the scope of this work, and referring to them respectively as $\alpha_{0,V}$, $\alpha_{0^\prime,V}$, $\alpha_{2,V}$ and $\alpha_{1,V}$, we expect
\begin{align}
    \frac{\alpha_{1,V}}{m_1} \sim \frac{\alpha_{0,V}}{m_0} \sim  \frac{\alpha_{0^\prime,V}}{m_{0^\prime}} \sim \frac{\alpha_{2,V}}{m_2}\,,
\end{align}
where $m_1$ is the mass of the $X(3872)$. The matrix element for the $X(3872)\to J/\psi V$ decay can be written as~\cite{Maiani:2004vq}
\begin{align} \label{eq:matrixX3872}
    \langle X(3872)|J/\psi V \rangle = \alpha_{1,V} \big(\bm{\epsilon}_X\times \bm{\epsilon}_{J/\psi}\big) \cdot \bm{\epsilon}_{V}\,.
\end{align}

The couplings $\alpha_{1,\rho}$ and $\alpha_{1,\omega}$ can be extracted from the branching ratios $\mathcal{B}(X(3872)\to J/\psi \pi\pi)\simeq 3.8\%$ and $\mathcal{B}(X(3872)\to J/\psi\pi\pi\pi)\simeq4.3\%$~\cite{ParticleDataGroup:2020ssz,Brazzi:2011fq,*Faccini:2012zv,*Albaladejo:2020tzt}. In particular, the partial widths for these decays can be computed as
\begin{align}
\begin{split}
    &\Gamma(J/\psi f) = \mathcal{B}(V\to f) \\ 
    & \quad \times\int_{s_\text{min}}^{s_\text{max}} ds \, \frac{\Gamma_{V\to f}(s)}{\Gamma_{V\to f}(m^2_V)} \, \text{BW}(s) \Gamma(J/\psi V)\,.
\end{split}
\label{GammaX}
\end{align}
Here $f=\pi\pi$ or $\pi\pi\pi$, $\mathcal{B}(V\to f)$ is the branching ratio for the decay of the light vector into the final state $f$, and $\Gamma(J/\psi V)$ is the decay width of the $X(3872)$ into $J/\psi$ and a light vector of invariant mass $s$, as computed from Eq.~\eqref{eq:matrixX3872}. Moreover, $s_\text{max}=(m_X-m_{J/\psi})^2$,  $s_\text{min}=(2m_\pi)^2$ if $V=\rho$ and $(3m_\pi)^2$ if $V=\omega$, and $\Gamma_{V\to f}$ are the decay rates reported in Appendix~\ref{App:PhSpace}. Using the Breit--Wigner width of the $X(3872)$ as recently measured by LHCb, $\Gamma_X = 1.39 \pm 0.34~\MeV$~\cite{LHCb:2020xds}, one finds
\begin{align}
    \alpha_{1,\rho}\simeq 342~\MeV\,, \quad \text{ and } \quad \alpha_{1,\omega} \simeq 1119~\MeV\,.
\end{align}

Diagrams with an intermediate $\rho$ and $\omega$ will now both contribute coherently to the total width in two photons. Starting again from Eqs.~\eqref{eq:matrixelements}, one finds, for the scalars and tensor mesons,
\begin{subequations}
\begin{align}
    \Gamma_{\gamma\gamma}^{X_{0}^{(\prime)}} &= \frac{\pi \alpha^2 \kappa_{\psi}^2 f_\psi^2}{m_{0^{(\prime)}}m_\psi^4}\left| \sum_{V=\omega,\rho} \frac{\kappa_V f_V \alpha_{0,V}}{m_V^2-i m_V \Gamma_V} \right|^2\,, \\
    \Gamma_{\gamma\gamma}^{X_2} & = \frac{7\pi\alpha^2\kappa_\psi^2 f_\psi^2}{30 m_2 m_\psi^4}\left| \sum_{V=\omega,\rho} \frac{\kappa_V f_V \alpha_{2,V}}{m_V^2 - i m_V\Gamma_V} \right|^2\,,
\end{align}
\end{subequations}
with $\Gamma_V$ the width of the light vector. Putting everything found so far together, one obtains the partial widths and production cross sections in UPCs reported in Table~\ref{tab:sigmas3872}.
\begin{table}[t]
    \centering
    \begin{tabular}{c||c|c}
        State, $J^{PC}$ & $\Gamma_{\gamma\gamma}$ (eV) & 
        $\sigma(\text{Pb}\,\text{Pb}\to\text{Pb}\,\text{Pb}\,X)$ (nb) \\
        \hline\hline
        $X_0(\sim 3770)$, $0^{++}$ & $\sim 6.3$ & $\sim 185$\\
        $X_0^\prime(\sim 4000)$, $0^{++}$ & $\sim 6.7$ & $\sim 156$ \\
        $X_2(\sim 4000)$, $2^{++}$ & $\sim 1.6$ & $\sim 187$\\
    \end{tabular}
    \caption{Partial widths in two photons and corresponding production cross sections in UPCs for the scalar and tensor elements of the $X(3872)$ multiplet. Again, the values are computed for $\sqrt{s_\text{NN}}=5.5~\TeV$.}
    \label{tab:sigmas3872}
\end{table}

Due to the small widths of $X(3872)\to J/\psi V$, the resulting production cross sections are smaller than that of the $X(6900)$. Nonetheless, they are still larger than that of the $X(3872)$ as observed produced promptly in $pp$ collisions~\cite{Artoisenet:2009wk}. Moreover, the decay of the exotic in its final state is dominated by the $S$-wave component just like for the $X(6900)$. For this reason, we expect similar distributions as in Figure~\ref{fig:dists}.


\section{Conclusion}

We proposed to look for compact tetraquarks in ultra-peripheral heavy ion collisions. In particular, we focus on those resonances whose observation represents a clear indication of a compact tetraquark nature. 

The first is the $X(6900)$, recently discovered by LHCb and having a $cc\bar c\bar c$ valence structure. Since there is no known mechanism that can bind together two charmonia in a loosely bound molecule, this state is likely compact. We find that, due to its strong coupling to a di-$J/\psi$ final state, this resonance is expected to be produced copiously in ultra-peripheral collisions. Its study in this context would allow to shed further light into its properties.

The other possible direction that would demonstrate the existence of four-quark objects in short distance QCD is the observation of a complete flavor-spin multiplet, very much analogously to what happened for standard mesons and baryons. In particular, beside the famous charged partners of the $X(3872)$, the missing pieces of the $S$-wave diquarkonia are the scalar and tensor states. These too are expected to be produced in ultra-peripheral collisions, with cross sections larger than the (large) prompt production cross section of the $X(3872)$ in proton-proton collision.

Ultra-peripheral heavy ion collisions are an ideal setup for different sorts of exotic searches, and they could provide a key insight into a yet unanswered question of strong interactions.


\begin{acknowledgments}
We are grateful to F.~Antinori, G.~M.~Innocenti and A.~Uras for very fruitful discussions and for encouraging the present study. A.E. is a Roger Dashen Member at the Institute for Advanced Study, whose work is also supported by the U.S. Department of Energy, Office of Science, Office of High Energy Physics under Award No. DE-SC0009988. A.E. has also received funding by the Swiss National Science Foundation under Contract No. 200020-169696 and through the National Center of Competence in Research  SwissMAP. 
The work of C.A.M. is supported by the Swiss National Science Foundation (PP00P2\_176884).
A.P. has received funding from the European Union's Horizon 2020 research and innovation program under the Marie Sk{\l}odowska-Curie Grant Agreement No.~754496.
\end{acknowledgments}

\vspace{10pt}
\appendix

\section{$\rho\to 2\pi$ and $\omega\to 3\pi$}\label{App:PhSpace}

The process $\rho\to \pi^+\pi^-$ is a $P$-wave decay. The matrix element is given by
\begin{align}
    \mathcal{M} = C_\rho  \eps_{\rho}^\mu (p_{+}-p_{-})_{\mu}\,, 
\end{align}
where $\eps_{\rho}^\mu$ is the polarization vector of the $\rho$ and $p_{+(-)}$ is the four momentum of the $\pi^{+(-)}$. For the decay rate, in the rest frame of the $\rho$, one finds
\begin{align}
    \Gamma_{\rho\to 2\pi}(s) \propto \int_{-1}^{1} d\cos\theta\,  \frac{p}{\sqrt{s}} \sum_\text{pol}\left|\mathcal{M}\right|^2\,,
\end{align}
where $s$ is the $\rho$ invariant mass, $\theta$ is the angle between the $\rho$ quantization axis and the $\pi^+$ direction of flight in the $\rho$ rest frame, and $p = \lambda^{1/2}\!\left(s, m_+^2, m_-^2\right)\big/ 2 \sqrt{s}$ with $m_{0,\pm}$ the pion masses. Performing the integral, we have
\begin{align}
    \Gamma_{\rho\to 2\pi}(s) \propto \frac{p^3}{\sqrt{s}}\,.
\end{align}

In the decay $\omega\to \pi^+\pi^-\pi^0$, the pair $\pi^+\pi^-$ has an angular momentum $\ell=1$ and a relative angular momentum $\ell=1$ with the $\pi^0$. The matrix element for this process is given by
\begin{align}
    \mathcal{M} = C_\omega \varepsilon_{\mu\nu\rho\sigma} P_{\omega}^\mu\eps_{\omega}^\nu p_+^\rho p_-^\sigma\,, 
\end{align}
and for the decay rate one finds
\begin{align}
    \begin{split}
        \Gamma_{\omega\to 3\pi}(s) &\propto \int_{(m_++m_-)^2}^{(\sqrt{s} - m_0)^2} d\sigma \int_{-1}^{1} d\cos\theta^+  \frac{\sqrt{\sigma} p}{s} \\
        & \quad \times \frac{q}{\sqrt{\sigma}} \sum_\text{pol} \left|\mathcal{M}\right|^2\,.
    \end{split}
\end{align}
where $\sigma$ is the $\pi^+\pi^-$ invariant mass, $s$ is the $\omega$ invariant mass and $\theta^+$ is the angle between $\omega$ and $\pi^+$ directions of flight in the $\pi^+\pi^-$ rest frame. In addition, $p = \lambda^{1/2}\!\left(s, m_0^2, \sigma\right)\big/ 2 \sqrt{\sigma}$ and $q = \lambda^{1/2}\!\left(\sigma, m_+^2, m_-^2\right)\big/ 2 \sqrt{\sigma}$.

Neglecting irrelevant constants which cancel in Eq.~\eqref{GammaX}, we find
\begin{align}
    \Gamma_{\omega\to 3\pi}(s) \propto \int_{(m_++m_-)^2}^{(\sqrt{s} - m_0)^2} d\sigma \frac{(\sqrt{\sigma} p)^3}{s} \frac{q^3}{\sqrt{\sigma}}\,.
\end{align}

\bibliography{bibliography}

\end{document}